\documentclass[useAMS,usenatbib,usegraphicx]{mn2e}

\title[Low-mass companion in the HL Tau disc]{Enhanced Dust 
Emission in the HL Tau Disc: A Low-Mass Companion in Formation?}
\author[J. S. Greaves et al.]{J. S. Greaves$^{1}$\thanks{E-mail: 
jsg5, at st-andrews.ac.uk}, A. M. S. Richards$^{2}$, W. K. M. 
Rice$^{3}$ and T. W. B. Muxlow$^{4}$\\
$^{1}$ Scottish Universities Physics Alliance, University of St 
Andrews, Physics \& Astronomy, North Haugh, St Andrews, Fife KY16 
9SS, UK. \\
$^{2}$Jodrell Bank Centre for Astrophysics, Turing Building, 
University of Manchester, Manchester M13 9PL, UK. \\
$^{3}$Scottish Universities Physics Alliance, Institute for 
Astronomy, University of Edinburgh, Blackford Hill, Edinburgh 
EH9 3HJ, UK. \\ 
$^{4}$Jodrell Bank Observatory, University of Manchester, 
Macclesfield, Cheshire SK11 9DI, UK. }

\begin{document}

\date{Accepted 2008. Received 2008; in original form 2007}

\pagerange{\pageref{firstpage}--\pageref{lastpage}} \pubyear{2008}

\maketitle

\label{firstpage}

\begin{abstract}

We have imaged the disc of the young star HL Tau using the VLA at
1.3~cm, with $0.08''$ resolution (as small as the orbit of
Jupiter). The disc is around half the stellar mass, assuming a
canonical gas-mass conversion from the measured mass in large
dust grains. A simulation shows that such discs are
gravitationally unstable, and can fragment at radii of a few tens
of AU to form planets. The VLA image shows a compact feature in
the disc at 65~AU radius (confirming the `nebulosity' of
\citet{welch}), which is interpreted as a localised surface
density enhancement representing a candidate proto-planet in its
earliest accretion phase. If correct, this is the first image of
a low-mass companion object seen together with the parent disc
material out of which it is forming. The object has an inferred
gas plus dust mass of $\approx 14$~M$_{\rm Jupiter}$, similar to
the mass of a proto-planet formed in the simulation. The disc 
instability may have been enhanced by a stellar flyby: the proper 
motion of the nearby star XZ Tau shows it could have recently 
passed the HL Tau disc as close as $\sim 600$~AU.

\end{abstract}


\section{Introduction}

The mechanisms by which giant planets form are uncertain. 
Core-accretion models \citep[e.g.]{pollack,hubickyj} have
successfully linked high abundances of rocky elements in the star
to higher planet-probability \citep[e.g.]{fischer,santos}, and
may explain planets with substantial rocky cores \citep{sato} --
but have theoretical difficulties with slow planetary cooling
that limits mass accretion rates and with rapid inwards migration
leading to loss of cores into the star. Also, the time of around
6~Myr to complete Jupiter may conflict with the infrared
detection rate of discs that declines close to zero by 6~Myr
\citep{haisch}, and with the latest ages of $\sim 15$~Myr when
gas is detected \citep{dent} since planet completion takes longer
in low-mass discs. The alternative model of gravitational
instability can create a proto-planet very rapidly, on dynamical
(orbital) timescales \citep{boss,rice05}, provided that the disc
cooling time is similarly short \citep{gammie01,rafikov}. Only
relatively rare discs of $\ga 0.1 \times M_{star}$ will be
unstable, but recent radio studies \citep[e.g]{rodmann} that
account for mass in large dust grains may boost more of the total
disc masses into this category.  Since metre-sized particles are
gathered up into the gas fragments \citep{rice06}, this model may
also account for solid cores to giant planets. 

\begin{figure*}
\label{fig1}
\includegraphics[width=145mm,angle=0]{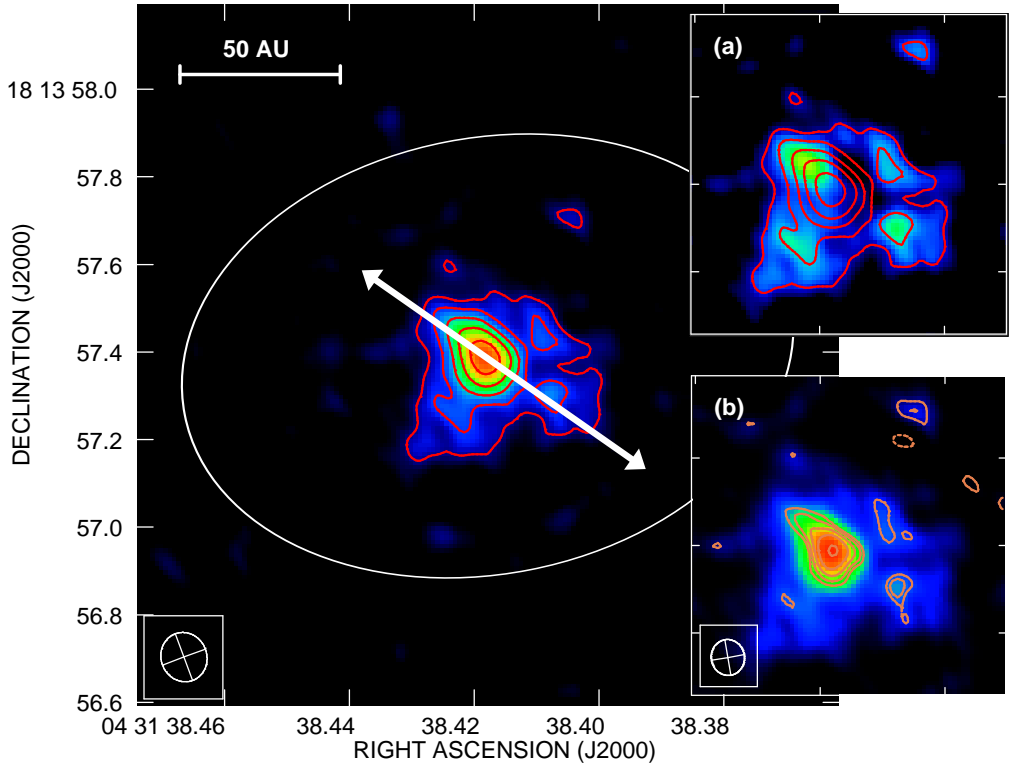}
\caption{VLA 1.3 cm images towards HL Tau. Main image: natural 
weighting with a beam (small inset) of $0.11''$ for maximum 
sensitivity; contours increase by $\sqrt{2}$ and are at 
(4.0, 5.7, 8.0, 11.4, 16.0) $\times 17 \umu$Jy/beam 
(1$\sigma$). The arrow indicates the jet axes and the ellipse 
shows the approximate extent of the inclined disc in a previous 
$0.6''$ resolution image at 2.7~mm \citep{looney}. The compact 
object lies to the upper-right. Upper inset: same but with 
central peak of 263 $\umu$Jy subtracted, to highlight the 
jet bases and two features at $\approx 20$~AU at the ends of 
the disc major axis. Contours from the unsubtracted image are 
overlaid. Lower inset: uniform weighted image at higher 
resolution of $0.08''$, with contours at (3.0, 4.2, 6.0, 8.5, 
12.0) $\times 21 \umu$Jy/beam (1$\sigma$), highlighting the 
compact object.
}
\end{figure*}

Imaging a planet within its birth-disc would illuminate the
processes involved. Companions of several Jupiter masses upwards
have been imaged, but large separations from the primary and in
some cases small mass-ratios of the two components suggest that
these objects may have formed like binary stars \citep{luhmann}. 
No discs within these systems have been imaged, so the formation
processes remain obscure -- the most direct observational
evidence for a forming object is a cleared cavity within the disc
of AB~Aur \citep{oppenheimer}. 

In this letter, we present the first candidate for a low mass
companion imaged in the accretion stage and within the parent
disc. Such an object would be expected to appear as a cool
condensation of dust and gas, possibly without a distinct dense
core as sedimentation timescales for large dust grains can be a
few $10^4$ years \citep{helled}. Here, we use radio-wavelength
data to trace the thermal emission from large dust particles in
the disc around HL Tau. This pre-main-sequence Class~I (remnant
envelope) object has been modelled by \citet{tom} at around
0.33~M$_{\odot}$ and 5~L$_{\odot}$, seen at $< 10^5$~years old.
The HL~Tau disc was selected as one of the brightest known at
millimetre wavelengths, with estimates for gas plus dust mass of
up to 0.1~M$_{\odot}$ \citep{beckwith}, and thus within the
disc-to-star mass regime where instability could occur.
Millimetre interferometry (resolving out the envelope) has shown
emission from the dust-disc extending out to at least $\sim
100$~AU radius \citep{wilner,mundy,lay,looney,rodmann}. 

\section{Observations}

Observations of HL Tau in a $10''$ field centred at RA
04:31:38.4034, Dec. 18:13:57.748 (J2000) were made with the Very
Large Array at 1.3~cm wavelength. In the largest A-configuration
plus the 50 km-distant Pie Town antenna, the VLA was sensitive to
scales down to $0.08''$, a factor of three higher than the best
previous resolution of $0.25''$ \citep{welch}. At the Taurus
distance of $\approx 140$~pc, this gives a resolution of just
over 10~AU, equivalent to the orbit of Jupiter.  The 22.5 GHz
data were obtained over two runs in 2006 Mar and Apr, with a
total usable time of $\approx 12$~hours. We used the phase
reference source 04311+20376, 2.4$^{\circ}$ from HL Tau and the
primary flux scale was provided by 3C286.  We also corrected the
antenna pointing and refined the amplitude calibration with the
aid of bright compact sources including 0552+398. We followed
standard VLA observational and data reduction procedures as
described at http://www.vla.nrao.edu/astro/ including procedures
for High Frequency Data Reduction. Natural weighting gave a beam
size of 114 by 104 mas and 1$\sigma$ sensitivity of 17 $\umu$Jy
per beam.  The data were also reconstructed with uniform
weighting, giving a smaller beam of 82 by 76 mas with 1$\sigma$
sensitivity of 21 $\umu$Jy/beam. Systematic positional
uncertainties are $\sim 30$~mas, less than the resolution. 

We also observed HL Tau at 5~cm wavelength with the MERLIN array
(using up to 6 antennas) in 2006 Jan-Feb for a usable total time
of 20 hr including calibration. We used the phase reference
source B0425+174 at 1$^{\circ}$ separation and followed
procedures described in \citet{diamond}. These were some of the
first observations made with six 5-GHz receivers (not all
cryogenic), reaching a 1$\sigma$ sensitivity of 100 $\mu$Jy
beam$^{-1}$ using a 100-mas restoring beam. The images were
sensitive to scales of 0.04--0.8 arcsec depending on weighting. 
We re-observed in 2007 Apr at 6-cm wavelength; the combined data
reached a sensitivity of 55 $\umu$Jy beam$^{-1}$.  This gives a
formal 3$\sigma$ upper limit for 5--6-cm emission on these
scales of 165 $\umu$Jy beam$^{-1}$.  No emission brighter than
100 $\mu$Jy beam$^{-1}$ was detected within 0.2 arcsec of the
compact object discussed in Section 3.

\section{Results}

The image (Figure 1) shows a bright inclined disc out to around
30 AU, i.e. similar to Neptune's orbit. The elliptical morphology
and similar orientation to lower-resolution images confirm that
this is emission from the dust disc, while the two orthogonal
extensions (upper inset) are presumably the bases of the bipolar
jet, previously seen at 1-2$''$ from the star
\citep{rodmann,looney}. Fainter dust emission is detected out to
100 AU, declining with radius as r$^{-1.5 \pm 0.5}$ (from an
error-weighted fit using elliptical annuli from $0.1-0.7''$, with
a correlation coefficient of 0.9).  Assuming emission weighted by
$r^{-1/2}$ for grains in thermal equilibrium with the star, the
disc surface density declines at $r^{-1 \pm 0.5}$, roughly like
the young outer Solar System \citep{davis}. Imaged features
include the extended disc ($\sim 1350$ $\umu$Jy); a central peak
of $\sim 300$~$\umu$Jy located at (J2000) 04 31 38.4184, +18 13
57.387 (2 mas fit errors); the extensions to the NE and SW
interpreted as jets, each of flux $\approx 100 \pm 25$~$\umu$Jy;
and a clump of $78 \pm 17$~$\umu$Jy offset from the central peak
by 380 mas at position angle (PA) of $30^{\circ}$ clockwise. The
jet, disc and clump features are highlighted in a
higher-resolution image (lower inset).

This compact clump is at a projected stellar separation of 55~AU,
or orbiting at around 65 AU if corrected for projection assuming
a disc inclination of $60^{\circ}$ \citep{wilner}. The clump is
here resolved for the first time. A `nebulosity' was previously
reported in a BIMA 1.4 mm image of the disc \citep{welch},
separated by 70 AU from the star at PA $-40^{\circ}$. Within the
BIMA resolution of $0.25''$ (35 AU), this is coincident with our
VLA 1.3 cm peak at 55 AU, $-30^{\circ}$. The two independent
detections give high confidence that this feature is real, while
it is seen here in detail for the first time. The earlier image
of an unresolved flux enhancement could have been attributed to
an ordinary disc asymmetry such as a large-scale perturbation,
but in our new data a clump is clearly seen. It is compact
(full-width half-peak sizes of $20 \pm 12$~AU by $\leq 12$~AU),
three times brighter than the local flux level of the disc, and
clearly separated (by five resolution elements) from the stellar
position. Given the similarity to proto-planets formed in
simulations (Figure~2), we propose this object as a candidate for
the earliest stage of growth of a low-mass companion. 

\begin{figure}
\label{fig2}
\includegraphics[width=85mm,angle=0]{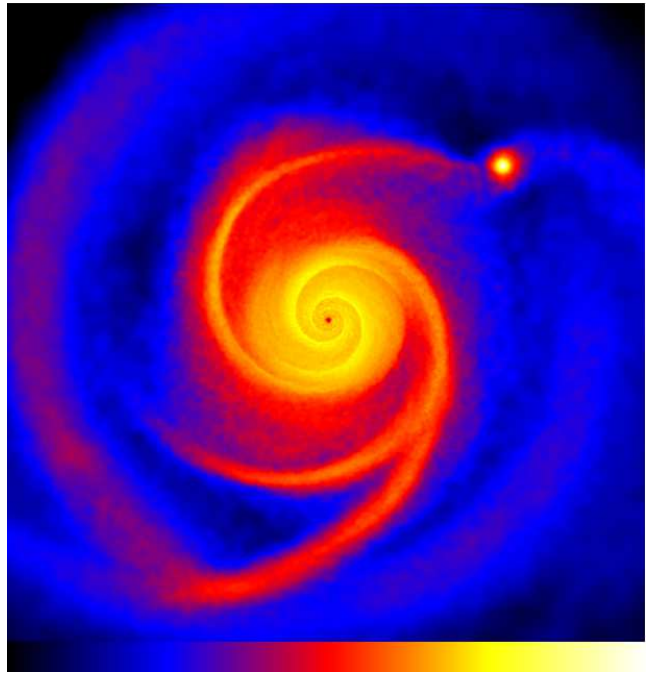}
\caption{Example image from an SPH simulation (see text) showing 
the surface density structure of a 0.3~M$_{\odot}$ disc around a 
0.5~M$_{\odot}$ star. A single dense clump has formed in the disc 
(upper right), at a radius of 75 AU and with a mass of $\approx 
8$~M$_{Jupiter}$. 
}
\end{figure}

\subsection{Robustness}

We tested whether this +4.5$\sigma$ feature could be an artefact.
The noise/pixel has a very Gaussian distribution across the
image, so the probability of a random fluctuation of $\geq 78
\umu$Jy/beam occurring within the projected disc area out to
100~AU should be only $\sim 0.03$~\%.  As a test, we added 33
fake `planets' to the visibility data, with the flux density of
the observed clump, scattered at $1-3''$ offsets, and followed
the standard imaging process. The recovered objects follow a
roughly Gaussian distribution, with mean and dispersion in flux
density of $75\pm25$~$\umu$Jy.  This distribution was compared to
the fluxes of 33 random positions, away from the disc, and no
noise pixels at these positions exceed 3$\sigma$.  The faintest
fake `planets', in the lower 1$\sigma$ tail of their flux
distribution, were close to the off-source 3$\sigma$ noise level,
showing that a true planet of such low intensity might be lost in
the noise -- but no false planet is seen emerging as a 3$\sigma$
artefact, let alone at the 4.5$\sigma$ level of the actual
feature in the disc. 

We then subtracted out the bright central emission (Figure 1, upper
inset) and found that the compact object is unaffected;
conversely we added a model of this emission at random positions
across the field -- no spurious features above 3$\sigma$ were seen
at the distance of our candidate. Hence, the clump is not an
artefact arising from imperfect deconvolution of the
interferometric image. Finally we split the data into two separate
frequency bands and alternatively into left and right circular
polarizations. The clump was always recovered at a similar
intensity, at between 2.5$\sigma$ and 4.8$\sigma$ significance
depending on the quality in the partial dataset, supporting a real
detection. 

We also investigated whether an unrelated source could be seen
through the HL Tau disc -- such background radio objects would
typically be active galactic nuclei. This is improbable as even
the most distant extragalactic faint radio sources are at least
$\sim 0.4''$ in diameter \citep{muxlow}, five times larger than
our beam. Conversely, a source would need to be well over 1~mJy
in lower-resolution surveys to be detectable, and as AGN emission
rises at long wavelengths (opposite to dust), this would be a
rare bright object. Extrapolating a 1~mJy 1.3-cm source with a
spectral index of $\alpha < -0.2$ (for 92~\% of radio sources
with multi-frequency data \citep{volmer} and flux $\propto
\nu^{-\alpha}$) yields a signal $> 2$~mJy at 20~cm.  The VLA
FIRST survey (http://sundog.stsci.edu/first/) shows such an
object would turn up in our $10'' \times 10''$ field in about 1
in 3000 cases (actually much lower since in this flux range most
sources are $\gg 0.4''$ in size). Moreover, such a source would
have probably appeared at 5-cm: for $\alpha < -0.2$ the
counterpart to the 1.3~cm feature would be of $\geq
100$~$\umu$Jy. The MERLIN image showed no features above 100
$\umu$Jy in this region (3$\sigma$ limit of 165 $\umu$Jy), so a
background synchrotron source is ruled out with high confidence.
A {\it millimetre} counterpart (see below) indicates a dust-like
spectrum for the clump -- a distant starburst galaxy could
potentially have such a spectral energy distribution, but these
are rare. Extrapolating from the 1.3~cm flux would lead to a
source of $> 20$~mJy at 0.85 mm wavelength, with a corresponding
probability of $< 10^{-6}$ \citep{coppin} within 100 AU of HL
Tau.

\begin{figure}
\label{fig3}
\includegraphics[width=80mm,height=52mm,angle=0]{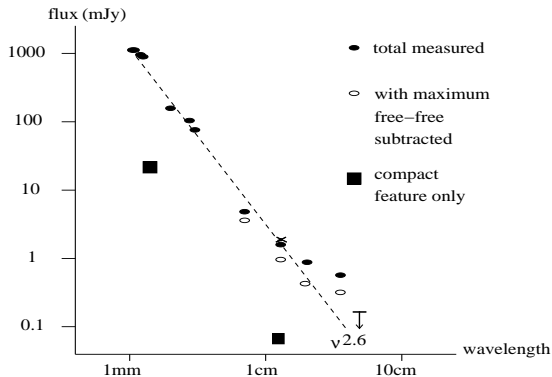}
\caption{Spectral energy distribution of clump (squares) and  
integrated flux around HL Tau, including VLA (cross), MERLIN 
upper limit (arrow) and results from \citet{rodmann} (filled 
ovals). Unfilled ovals show the {\it minimum} flux-contribution 
from dust, after subtracting free-free flux extrapolated from 
the MERLIN upper limit assuming $F_{\nu} \propto \nu^1$, the 
steepest slope typically seen for ionised stellar winds 
\citep{anglada}. A  flatter ($\nu^{-0.1}$) subtraction gives an 
improved $\nu^{2.5}$ dust power-law fit. 
}
\end{figure}

\subsection{Measurements}

The spectral energy distribution confirms that circumstellar dust
is detected.  The integrated centimetre flux was previously
thought to be from the ionised stellar wind, but Figure~3 shows
dust emission to $\lambda \geq 3.6$~cm. The dust spectrum of
F$_{\nu} \propto \nu^{2.5-2.6}$ is characteristic of emission
from a population of particles extending in size up to at least
three times the observing wavelengths \citep{draine}, and hence
here to bodies of $> 10$~cm. The spectrum of the condensation is
$\propto \nu^{2.5}$ from the fluxes of $23 \pm 5$ mJy at 1.4 mm
and $78 \pm 17$ $\umu$Jy at 1.3 cm (neglecting lower-resolution
data noted by \citet{welch} as surrounding disc flux may be
included), again implying that very large particles are present.
This would agree with simulations \citep{rice06} in which
`boulder'-like bodies of around metre-size are most readily
captured in unstable regions. 

At 1.3~cm, the measured disc flux excluding the jets and region
inside 5~AU is $\sim 1350$~$\umu$Jy. Assuming simplistically that
the dust particles are in thermal equilibrium with the star, we
adopt the best-fit 5~$L_{\odot}$ from \citet{tom}, giving 90~K at
a characteristic disc radius of $\sim 20$~AU (Figure 1, upper
inset) and 50~K at the clump orbit. For an opacity of 7$\times
10^{-4}$~cm$^2$/g at 1.3 cm (for populations extending up to 1-10
cm particles \citep{draine}, and assuming that all the original
gas and dust are present so that a canonical gas-to-dust mass
ratio of 100 applies), the disc contains around 0.13~M$_{\odot}$
in total. This exceeds previous estimates of up to
0.1~M$_{\odot}$, due to the additional large dust and consequent
scaling up of the total mass. The clump comprises $\approx
14$~M$_{\rm Jupiter}$, with the uncertainty dominated by the
scaling with temperature; the errors in flux and distance
contribute at up to $\sim 30$~\% levels while adopting a
different opacity (e.g. $\nu^{0.5}$ extrapolation from a standard
0.01~cm$^2$/g at 1~mm) reduces the mass by factor of two. 
The central peak may also contain a mass reservoir -- the 1.3 to
6~cm spectrum is just consistent with a wind origin, but allows
up to $\sim 13$~M$_{\rm Jupiter}$ of gas and dust to be present
if the 1.3~cm flux is dust-dominated. This is similar to the
primordial 12~M$_{\rm Jupiter}$ out to Jupiter's orbit
\citep{davis}, and so planets might form here by core-accretion;
if large grains are present this may also solve the planetary
radiative cooling problem by reducing the dust opacity
\citep{hubickyj}. 

\subsection{Simulations}

\begin{figure}
\label{fig4}
\includegraphics[width=78mm,angle=0]{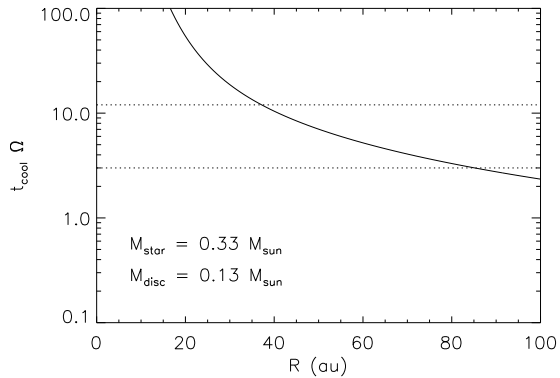}
\caption{Plot of cooling time as a function of radius for a disc 
mass of 0.13~M$_{\odot}$ around a star of mass 0.33~M$_{\odot}$. 
The disc surface density declines as $r^{-1.5}$ to an outer 
radius of 100~AU, and the instability parameter is $Q \sim 1$.
The quantity $t_{\rm cool} \Omega$ is a dimensionless quantity 
relating the cooling time to the local orbital period; for  
$t_{\rm cool} \Omega = 2 \pi$ the cooling time equals  
the orbital period. The dotted lines show critical 
values below which fragmentation occurs for different equations 
of state \citep{rice05}. 
}
\end{figure}

We ran an example 250,000-particle Smoothed Particle
Hydrodynamics (SPH) simulation to investigate if the HL Tau disc
could be gravitationally unstable (Figure~2). The disc is assumed
to have an initial surface density profile of $\Sigma \propto
r^{-1.5}$ and initial outer radius of 100 AU. The simulation
evolves under a radiative transfer formalism \citep{stam} in
which each particle is allowed to cool towards its equilibrium
temperature, determined using the local optical depth (dependent
on material opacity, here taken to be half-Solar). The disc is
allowed to heat up through $p dV$ work and viscous dissipation.
Relative to the local dynamical timescales, the cooling times are
slow in the inner disc where the optical depth is high and faster
in the outer disc.  Fragmentation is expected if the cooling time
is less than a few orbital periods, and depends on the equation
of state \citep{rice05}. For the estimated star and disc masses
of HL Tau, Figure 4 shows that the cooling time should be fast
enough for fragmentation beyond $\approx 40$~AU. In the
simulation, we adopted somewhat higher star and disc masses (but
within the uncertainties) and allowed the disc to cool as low as
10~K, which favours fragmentation (Figure~2). A single dense
clump has formed in the disc, and is located at a radius of 75 AU
and has a mass of $\sim 8$~M$_{\rm Jupiter}$, although it may
continue to accrete from the disc. These properties are similar
to those of the actual observed object around HL Tau. The
simulation results will vary with the chosen opacity -- a smaller
value produces additional clumps, while a larger one could
inhibit fragmentation altogether -- and with the surface density
profile -- a flatter profile yields more outer-disc mass 
and higher tendency to fragment. 

\section{Discussion}

Including the large grains now detected, the HL~Tau disc mass is
$\approx 0.13$~M$_{\odot}$. \citet{tom} find good fits to the
stellar mass for 0.2--1~M$_{\odot}$; for their best-fit value of
0.33~M$_{\odot}$ the disc is around $0.4 M_{\rm star}$. This
proportionally massive disc should be gravitationally unstable,
and a simulation at the higher end of the $M_{\rm disc, star}$
ranges confirms that planetary objects could form at a few tens
of AU. The VLA data show such a flux peak in the parent disc
material, interpreted here as a surface density enhancement. 
(The clump is three times brighter than the local disc flux,
while warming of the gas by gravitational collapse should only
contribute marginally to higher emission; simulation results
suggest the beam-averaged dust temperature is raised by $\approx
50$~\%.) This clump at 65 AU from HL Tau lies in the appropriate
unstable region, and is compact as expected for a low-mass object
accreting from the disc. 

The simulated disc is unstable for the adopted parameters, but
external forces could have increased the real disc's tendency to
fragment. Notably, another cluster member, XZ Tau, appears
close-by, which is unusual within the diffuse Taurus association, 
and
the relative motions suggest a possible recent encounter of the
two stars.  Their line-of-sight distances are unknown, but the
similar radial velocities \citep{folha} suggest they are not
located in very different parts of the association, and in 2-D
the stars are presently diverging. XZ Tau lies $23''$ east of HL
Tau and the proper motions \citep{ducourant} are (+11,--19) and
(--3,--21) milliarcsec/year respectively (errors of 2-5
mas/yr). Around 1600 years earlier, the stars could thus 
have passed within $\sim 600$~AU (in 2-D projection).  Such an
event would have been dynamically recent, given that the compact
object has an orbital period of 900~years for $M_{star}$ of
0.33~$M_{\odot}$. 

The final mass of this still-forming companion may increase, by
absorbing more of the disc, but our estimate of 14~M$_{\rm
Jupiter}$ for the condensation is well down into the sub-stellar
regime. If all this material is accreted, the final object would
be around the brown dwarf / planet boundary by the definition of
short-lived deuterium-burning capability, which occurs at $\ga
12$--13~M$_{\rm Jupiter}$. A more recently developed definition
of a planet is a low-mass object that formed in the disc of a
star. This `origins' definition sidesteps the deuterium-burning
issue, which as \citet{chabrier} point out is irrelevant for the
evolution of brown dwarfs. In the case of HL~Tau `b', imaging the
object within the parent disc marks it as a candiate proto-planet
by this origins definition.

\section*{Acknowledgments}

The VLA is part of the NRAO, a facility of the National Science
Foundation operated by Associated Universities, Inc. MERLIN is a
National Facility operated by the University of Manchester at
Jodrell Bank Observatory on behalf of STFC.

\bsp

\label{lastpage}

\end{document}